\documentclass[twocolumn, onecolappendix]{aastex701}
\usepackage{amsmath}
\usepackage{enumitem}
\usepackage{newtxtext,newtxmath}
\usepackage{bm}

\hypersetup{breaklinks,colorlinks,citecolor=blue,urlcolor=blue,linkcolor=red}

\newcommand{\uatentry}[2]{\href{http://astrothesaurus.org/uat/#2}{#1 (#2)}}
\newcommand{\Msun}{\,{\rm M}_\odot}

\defcitealias{chen_starstream_2025-1}{C25}

\graphicspath{{./}{figures/}}

\shorttitle{\it \textit{StarStream} on \textit{Gaia}: Stream discovery and mass loss rate of globular clusters}
\shortauthors{\it Chen et al.}

\begin{document}

\title{\vspace{-6mm}\large \textit{StarStream} on \textit{Gaia}: Stream discovery and mass loss rate of globular clusters}

\author[0000-0002-5970-2563]{Yingtian Chen}
\affiliation{Department of Astronomy, University of Michigan, Ann Arbor, MI 48109, USA}
\email{ybchen@umich.edu}

\author[0000-0001-9852-9954]{Oleg Y. Gnedin}
\affiliation{Department of Astronomy, University of Michigan, Ann Arbor, MI 48109, USA}
\email{ognedin@umich.edu}

\author[0000-0003-0872-7098]{Adrian M. Price-Whelan}
\affiliation{Center for Computational Astrophysics, Flatiron Institute, New York, NY 10010, USA}
\email{aprice-whelan@flatironinstitute.org}

\correspondingauthor{Yingtian Chen}
\email{ybchen@umich.edu}

\begin{abstract}\noindent
We apply the automatic stellar stream detection algorithm \textit{StarStream} to \textit{Gaia} Data Release 3 and identify 87 stellar streams associated with Galactic globular clusters (GCs), including 34 high-quality cases with median completeness and purity both exceeding 50\%, as estimated from modeling mock streams. These detections double the number of known GC streams, and increase the fraction of GCs with tidal streams at high Galactic latitudes ($|b|>30^\circ$) to 75\%. In contrast to visual expectations, many new streams are wide or short, or misaligned with their progenitors' orbits. Taking advantage of the unbiased density measurements enabled by our method, we also estimate the mass loss rate for the progenitor GCs. We find that several low-mass, large-size clusters have enhanced mass loss rates, indicating that they are approaching complete tidal disruption.
\end{abstract}


\section{Introduction}

The advent of the \textit{Gaia} mission \citep{gaia_collaboration_gaia_2016}, in particular the inclusion of high-precision proper motions down to $G \approx 20$ by Data Release 2 \citep[DR2,][]{gaia_collaboration_gaia_2018}, has greatly reshaped our understanding of the Milky Way (MW) substructure by providing an all-sky map of stars in the full six-dimensional (6D) phase space.

By applying variations of the matched-filter technique \citep[e.g.,][]{rockosi_matched-filter_2002,grillmair_four_2009,bernard_serendipitous_2014,shipp_stellar_2018} to both the color--magnitude diagram (CMD) and proper motion space, astronomers have identified over one hundred thin, elongated structures recognized as stellar streams \citep[see the review by][]{bonaca_stellar_2025}. Notably, \citet{ibata_charting_2024} employed the \texttt{STREAMFINDER} algorithm \citep{malhan_streamfinder_2018} on \textit{Gaia} DR3 \citep{gaia_collaboration_gaia_2023} to uncover 87 thin streams, while \citet{hallin_via_2025} combined the \texttt{Via Machinae} method \citep{shih_via_2021,shih_via_2024} with the \texttt{Cathode} algorithm \citep{hallin_classifying_2022} to detect around 80 thin streams in \textit{Gaia} DR2.

Many stellar streams are elongated debris of tidally disrupted globular clusters (GCs, see \citealt{lynden-bell_ghostly_1995}). Their morphology and kinematics preserve rich information about interactions with the Galactic environment, including the dark matter halo \citep{koposov_constraining_2010,kupper_globular_2015,nibauer_galactic_2025}, the rotating bar \citep{pearson_gaps_2017,bonaca_variations_2020}, the tilting disk \citep{nibauer_slant_2024}, and encounters with subhalos, other GCs, or giant molecular clouds \citep{carlberg_pal_2012,ngan_using_2014,erkal_number_2016,erkal_sharper_2017,banik_probing_2018}.

In addition, stellar streams encode key properties of their progenitor GCs, such as the mass loss rate \citep[e.g.,][]{kupper_globular_2015,gieles_supra-massive_2021,chen_stellar_2025}, which is directly related to the stream density. Such measurements provide valuable constraints on the N-body simulations of dynamical evolution of GCs. However, most stream detections do not include quantified estimates of completeness and purity, leading to unknown systematic uncertainties in the inferred mass loss rates. Furthermore, fewer than 20 GCs have so far been confidently associated with streams, while most clusters \citep[$>150$,][]{hilker_galactic_2019} still lack stream detections. Although tidally stripped stars have been widely observed around GCs \citep[e.g.,][]{kuzma_constructing_2025}, thin, extended ``stream-like'' features remain absent along the orbits of most GCs.

Recent advances in stream formation theory provide insights to the missing-stream puzzle. \citet{amorisco_feathers_2015} pointed out that streams may appear dynamically hot or spatially complex, depending on the progenitor's mass and orbit. Moreover, streams can deviate the progenitor's orbit in a non-spherical or time-evolving Galactic potential \citep{sanders_streamorbit_2013,sanders_streamorbit_2013-1,panithanpaisal_breaking_2025}. These effects may be amplified depending on the viewing angle. As a result, traditional detection approaches based on the visual expectation that streams are thin features elongated along the progenitor's orbit tend to miss these ``irregular'' streams.

The limitations of traditional methods motivated us to develop \textit{StarStream} \citep[][hereafter \citetalias{chen_starstream_2025-1}]{chen_starstream_2025-1}, a physics-based method that makes no prior assumptions about stream morphology. The method employs kernel density estimation (KDE) to build a mixture model of stream and background stars, and incorporates the fast and accurate particle spray algorithm of \citet{chen_improved_2025} to generate a realistic stream model in the spatial, velocity, and color--magnitude spaces. \citetalias{chen_starstream_2025-1} quantified the detection performance of \textit{StarStream} using a suite of validation tests on a mock dataset tailored to \textit{Gaia} DR3. \textit{StarStream} demonstrates purity and completeness both above $\sim65\%$ at high Galactic latitudes ($|b|>30^\circ$), even after accounting for dust extinction.

The high detection quality makes \textit{StarStream} a powerful tool to uncover GC streams that may have been missed by previous methods. Its quantified performance further allows us to derive unbiased estimates of the mass loss rate for these clusters. In this work, we apply \textit{StarStream} to \textit{Gaia} DR3 fields around MW GCs. In \S\ref{sec:method}, we provide an overview of \textit{StarStream} and the \textit{Gaia} DR3 dataset. We then present the discovery of new streams in \S\ref{sec:detecting_streams}, followed by the calculation of the mass loss rates in \S\ref{sec:mass_loss}. Finally, we summarize and discuss our findings in \S\ref{sec:summary}.

\section{Method}
\label{sec:method}

We apply the \textit{StarStream} algorithm to \textit{Gaia} DR3 stars around MW GCs to identify potential stream members. In this section, we provide an overview of the \textit{StarStream} algorithm, including our adjustment to the \textit{Gaia} DR3 dataset.

\subsection{Overview of \textit{StarStream}}
\label{sec:StarStream}

We refer to \citetalias{chen_starstream_2025-1} for the complete description and validation of the \textit{StarStream} algorithm\footnote{The \textit{StarStream} algorithm is published on GitHub at \url{https://github.com/ybillchen/StarStream}, where we also provide example Python notebooks for running the code.}. Here, we briefly recap the key concepts of \textit{StarStream}.

\textit{StarStream} uses mixture modeling to distinguish the GC stream from the background. The probability density function is given by
\begin{equation*}
    p({\bm x}) = f_{\rm s}\, p_{\rm s}({\bm x}) + (1-f_{\rm s})\, p_{\rm bg}({\bm x})
\end{equation*}
where $\bm x$ is an arbitrary point in the multi-dimensional space of observables. We introduce the stream fraction $f_{\rm s}$ to characterize the ratio between the stream model $p_{\rm s}({\bm x})$ and background model $p_{\rm bg}({\bm x})$. Both models are represented by Gaussian KDE constructed on tracer particles in the multi-dimensional space. For \textit{Gaia} DR3 specifically, we use six observables, including two sky coordinates, two corresponding proper motions, $\rm BP-RP$ color, and $G$-band magnitude. Instead of directly using the raw right ascension and declination provided by \textit{Gaia}, we rotate the coordinate system for each GC such that the GC is located at the origin with the velocity vector along the new longitude. This coordinate system ensures an almost identity metric tensor ${\bf g}\approx{\bf I}$ around the GC, which is necessary for the KDE to yield the correct probability density.

To construct the stream model, we simulate mock stream for each GC using the particle spray model by \citet{chen_improved_2025}. This method requires integrating the orbits of both the progenitor and tracer particles in a predefined Galactic potential model. We employ the \texttt{MilkyWayPotential2022} model implemented within \texttt{gala} \citep{price-whelan_gala_2017,price-whelan_adrngala_2024}, which has been validated against MW mass measurements out to $\sim150$~kpc \citep{hunt_milky_2025}.

We release tracer particles over the last 1~Gyr assuming a uniform ejection rate $\dot{N}_{\rm tracer}=4\ {\rm Myr}^{-1}$. We then sample the mass of each particle from the \citet{kroupa_variation_2001} initial mass function, with the minimum stellar mass being set to the lowest possible mass of the closest tracer particle that remains above the detection limit. Based on the stellar mass, we calculate the color and magnitude of each tracer particle from the MESA Isochrones and Stellar Tracks \citep[MIST,][]{dotter_mesa_2016,mestric_exploring_2022} model. $\dot{N}_{\rm tracer}$ is a numerical parameter that controls only the number of tracers to sample position space, proper motion space, and the CMD. In \citetalias{chen_starstream_2025-1}, we verified that our adopted value is sufficient to fully sample these spaces.

The KDE is constructed on these tracers. The kernel is a product of one-dimensional kernels with the bandwidth adapted to 0.1 times the standard deviation of all tracer particles in each dimension, except that we fix the bandwidth to 0.1 for magnitudes and 0.02 for colors. Varying these values by a factor of $0.5-2$ has a negligible effect on our results. It is worth noting that we convolve the kernels with the observational uncertainties when evaluating the KDE for observed stars. This is equivalent to increasing the bandwidths of KDE kernels. Our KDE is constructed in the full multidimensional space of spatial coordinates, proper motions, colors, and magnitudes, as illustrated in Fig.~1 of \citetalias{chen_starstream_2025-1}. Therefore, our stream model naturally captures the curved structure of streams across all of these dimensions. 

For the background, we randomly select $10^4$ stars from the real data as the tracer particles for constructing KDE. We adopt bandwidths of $0.5^\circ$ for position, $1\ {\rm mas\,yr^{-1}}$ for proper motions, and $0.1$ for both color and magnitude. To speed up KDE evaluation, we employ the grid interpolation technique, where the grid spacings are set equal to the corresponding bandwidths. We have tested that the final results are not sensitive to the choice of bandwidths or grid spacing.

Once we have constructed the stream model and background model, we estimate the stream fraction $f_{\rm s}$ by maximizing the log-likelihood for $N$ stars in the observational dataset,
\begin{equation*}
    \ln{\cal L} \equiv \sum_{i=1}^N \ln\left[f_{\rm s} p_{\rm s}({\bm x}_i) + (1-f_{\rm s}) p_{\rm bg}({\bm x}_i)\right]
\end{equation*}
in which ${\bm x}_i$ represent the multi-dimensional coordinate of the $i$-th star in the dataset. The best-fit stream and background probability densities for star $i$ are then given by $f_{\rm s}p_{\rm s}({\bm x}_i)$ and $(1 - f_{\rm s})p_{\rm bg}({\bm x}_i)$, respectively. Therefore, we can define the membership probability that star $i$ belongs to the stream as
\begin{equation*}
    P_{{\rm s},i}\equiv\frac{f_{\rm s}p_{\rm s}({\bm x}_i)}{f_{\rm s}p_{\rm s}({\bm x}_i)+(1-f_{\rm s})p_{\rm bg}({\bm x}_i) }.
\end{equation*}
We consider stars with membership probability greater than a threshold $P_{\rm th}\equiv0.5$ to be identified as stream members.

Although this method works for most streams, it assumes that there is no significant overlap between multiple streams. An exception was noted by \citet{yuan_low-mass_2020} for the GC pair NGC 5024 and 5053, which reside near the Galactic pole and are separated by only $\sim1^\circ$. We therefore apply a special treatment using a three-component mixture model, as described in Appendix~\ref{sec:NGC_5024_5053}.

\subsection{Observational datasets}
\label{sec:obs}

We apply the same selection criteria as \citetalias{chen_starstream_2025-1} to \textit{Gaia} DR3. Specifically, we restrict to stars with valid measurements of right ascension, declination, proper motions, $\rm BP-RP$ color, $G$-band magnitude, and the corresponding uncertainties. We then select all stars with $G<20$ inside the $10^\circ$ cone around each GC. In addition, we employ an initial CMD cut around the extinction-corrected isochrone, with a tolerance offset $\Delta({\rm BP-RP}) = \pm0.5$ around the main sequence and the red-giant branch. We also extend the isochrone with $\Delta G = 1.5$ above the tip of the red-giant branch and around the horizontal branch to include stars clustered in those regions. These criteria select between 1 million stars (near the Galactic pole) and 30 million stars (near the Galactic center) around each GC.

To search for the stream around each GC, we need the mass, 3D positions, and 3D velocities of the progenitor GC to simulate the stream using our particle spray method. We use the values from the fourth edition of the \citet{hilker_galactic_2019} catalog\footnote{\url{https://people.smp.uq.edu.au/HolgerBaumgardt/globular/}}, which contains 162 MW GCs with valid measurements. 

To obtain the isochrone of each GC in the CMD, we use the MIST model, which requires the age, metallicity, and color excess $E(B-V)$ due to extinction assuming the \citet{cardelli_relationship_1989} reddening law. We use $E(B-V)$ from the measurements by D. Massari et al.\footnote{Private comm. in the context of the Cluster Ages to Reconstruct the Milky Way Assembly (CARMA) project \citep{massari_cluster_2023}.}, or from the 2010 edition of the catalog by \citet{harris_catalog_1996}\footnote{\url{https://physics.mcmaster.ca/~harris/mwgc.dat}} for GCs not included in the former. For the remaining streams without reliable extinction measurements, we use the extinction map by \citet{schlegel_maps_1998} recalibrated by \citet{schlafly_measuring_2011}. The three sources of $E(B-V)$ typically differ only by $\lesssim30\%$ for the overlapping GCs. To avoid extreme extinction that affects our detection quality, we exclude $\sim 20\%$ GCs with $E(B-V)>1$, leaving 128 GCs in our final sample.

We then fit the age and metallicity of individual GCs using a grid of age $=7$, $9$, $11$, $13$~Gyr and ${\rm [Fe/H]}$ varying between $\pm 1$ around the value in the \citet{harris_catalog_1996} catalog, with a uniform spacing of $0.1$. We use the age and metallicity that minimize the residual sum of squares of $\rm BP-RP$ color around the isochrone. We select GC member stars within the half-mass radius from the \citet{hilker_galactic_2019} catalog, with an additional proper motion selection $\Delta\mu_\alpha$, $\Delta\mu_\delta=\pm1~{\rm mas\,yr^{-1}}$. We have verified that the best-fit metallicities are consistent with the \citet{harris_catalog_1996} values, with a standard deviation of 0.4. Although multiple metallicity values may lead to similar fits due to the age--metallicity degeneracy, it is worth noting that the specific values do not significantly affect the performance of \textit{StarStream}, as long as we can reproduce the shape of the isochrone. For GCs with too few ($<10$) selected stars for reliable fit, we directly adopt metallicities from the \citet{harris_catalog_1996} catalog and a fixed age $=10$~Gyr. However, this catalog does not cover all GCs in our sample. For GCs without metallicity measurements, we assume $\rm  [Fe/H]=-1$. As we show in \citetalias{chen_starstream_2025-1}, such an approximation has negligible impact on the detection quality.

\section{Detecting streams in the Milky Way}
\label{sec:detecting_streams}

\subsection{Method validation}
\label{sec:method_validation}

\begin{figure*}
    \centering
    \includegraphics[width=0.9\linewidth]{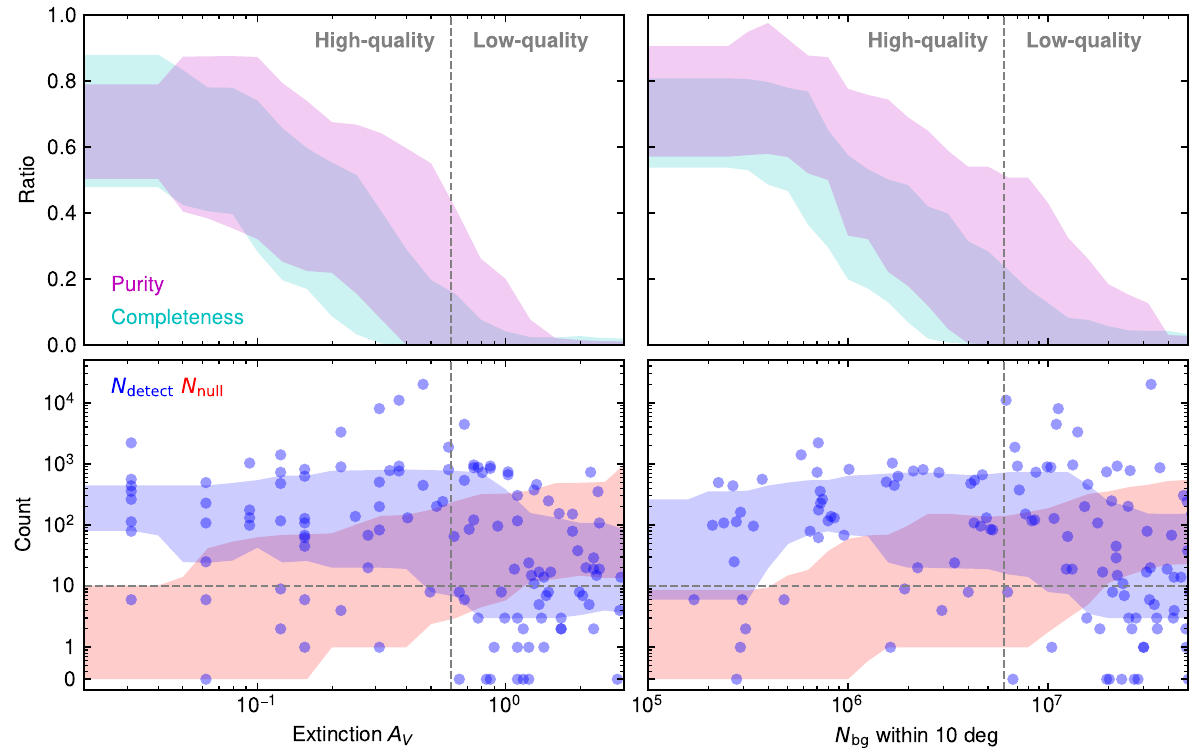}
    \caption{Detection quality metrics of \textit{StarStream} by \citetalias{chen_starstream_2025-1}. 
    \textbf{\textit{Upper row}}: Purity (magenta) and completeness (cyan) as functions of the progenitor's extinction $A_V$ (\textit{left}) and background density as characterized by $N_{\rm bg}$ within the $10^\circ$ search radius (\textit{right}). 
    \textbf{\textit{Lower row}}: Number of detections in the null test ($N_{\rm null}$, red) as a function of $A_V$ and background density. We also show the number of actual detection when applying \textit{StarStream} to MW GCs as blue lines, with individual detections shown as circles.
    Shaded regions represent the 25\%--75\% ranges, smoothed by a Gaussian kernel with bandwidth $=0.2$~dex for $A_V$ and $0.4$~dex for $N_{\rm bg}$. We show our threshold for high-quality detection, $A_V<0.6$ and $N_{\rm bg}<6\times10^6$, as vertical dashed lines. We also show the horizontal line to indicate the minimum selection threshold $N_{\rm detect}=10$.\vspace{2mm}}
    \label{fig:mock_real}
\end{figure*}

Before we report our discovery results, it is necessary to compare our detection with validation tests to rule out unreliable detections. We conducted these tests in \citetalias{chen_starstream_2025-1}, where we obtained the completeness and purity for individual mock streams from the \citet{holm-hansen_catalog_2025} catalog. We define completeness as the fraction of real stream members detected by \textit{StarStream}, and purity as the fraction of correctly detected members out of all detections. We also conducted a null test where we removed mock stream stars but applied the same \textit{StarStream} method to the region that used to have streams. Since we specifically tailored the mock dataset for \textit{Gaia} DR3, it is appropriate to compare our detections with these tests.

We note that these metrics strongly depend on the progenitor's extinction $A_V$ and background density. The latter is characterized by the number of background stars $N_{\rm bg}$ (accounting for the CMD selection in \S\ref{sec:obs}) within the $10^\circ$ search radius. In the \textit{upper row} of Fig.~\ref{fig:mock_real}, we show the completeness and purity as functions of these metrics. The values are directly taken from the with-extinction case of \citetalias{chen_starstream_2025-1} (see \S3.7 therein), excluding extremely high extinction cases following \S\ref{sec:obs}. As expected, both completeness and purity decrease with $A_V$ and $N_{\rm bg}$. They become less than 20\% at $A_V>0.6$ or $N_{\rm bg}>6\times10^6$. These values approximately correspond to the low-Galactic latitude region $|b|\lesssim20^\circ$, where the high dust reddening and background contamination significantly affect the detection quality. On the other hand, the method achieves acceptable completeness and purity for the low-$A_V$ and low-$N_{\rm bg}$ GCs. For mock GCs with $A_V<0.6$ and $N_{\rm bg}<6\times10^6$, the median completeness and purity are 50\% and 59\%, respectively. These values can even grow to $60\%-80\%$ near the Galactic poles. 

In the \textit{lower row} of Fig.~\ref{fig:mock_real}, we show the number of detections in the null test, $N_{\rm null}$. For comparison, we also show the actual number of detections $N_{\rm detect}$ when applying \textit{StarStream} to MW GCs in \textit{Gaia} DR3. For GCs with $A_V>0.6$ or $N_{\rm bg}>6\times10^6$, $N_{\rm detect}$ is indistinguishable from $N_{\rm null}$. However, in the high-latitude regions with $A_V<0.6$ and $N_{\rm bg}<6\times10^6$, the true detection number becomes significantly higher than $N_{\rm null}$ by up to two orders of magnitude. This is a strong evidence that we successfully detect streams around most GCs in this region. 

Based on the validation, we define high-latitude detections with $A_V<0.6$ and $N_{\rm bg}<6\times10^6$ as the \textit{high-quality} sample. We are confident that most detections in this sample are true detections with completeness and purity both above 50\%. Since GCs in this sample are at high latitudes $|b|\gtrsim20^\circ$, even their $10^\circ$ search radii do not intersect the high-$A_V$ Galactic plane, leading to a near-uniform distinction distribution. On the other hand, we define low-latitude detections with either $A_V\geq0.6$ or $N_{\rm bg}\geq6\times10^6$ as the \textit{low-quality} sample. Although we are less confident in this sample, it is still possible that they may be real. Since $N_{\rm null}$ varies between $0-10$ even in the low-$A_V$ and low-$N_{\rm bg}$ regions, we exclude streams with $N_{\rm null}\leq10$ from both samples to avoid false detections. Such criteria lead to a total of 87 GC streams, with 34 in the high-quality sample and 53 in the low-quality sample. These numbers greatly improve our knowledge of GC streams since less than 20 have been confirmed prior to this work.

\subsection{New streams}

\begin{figure*}
    \centering
    \includegraphics[width=\linewidth]{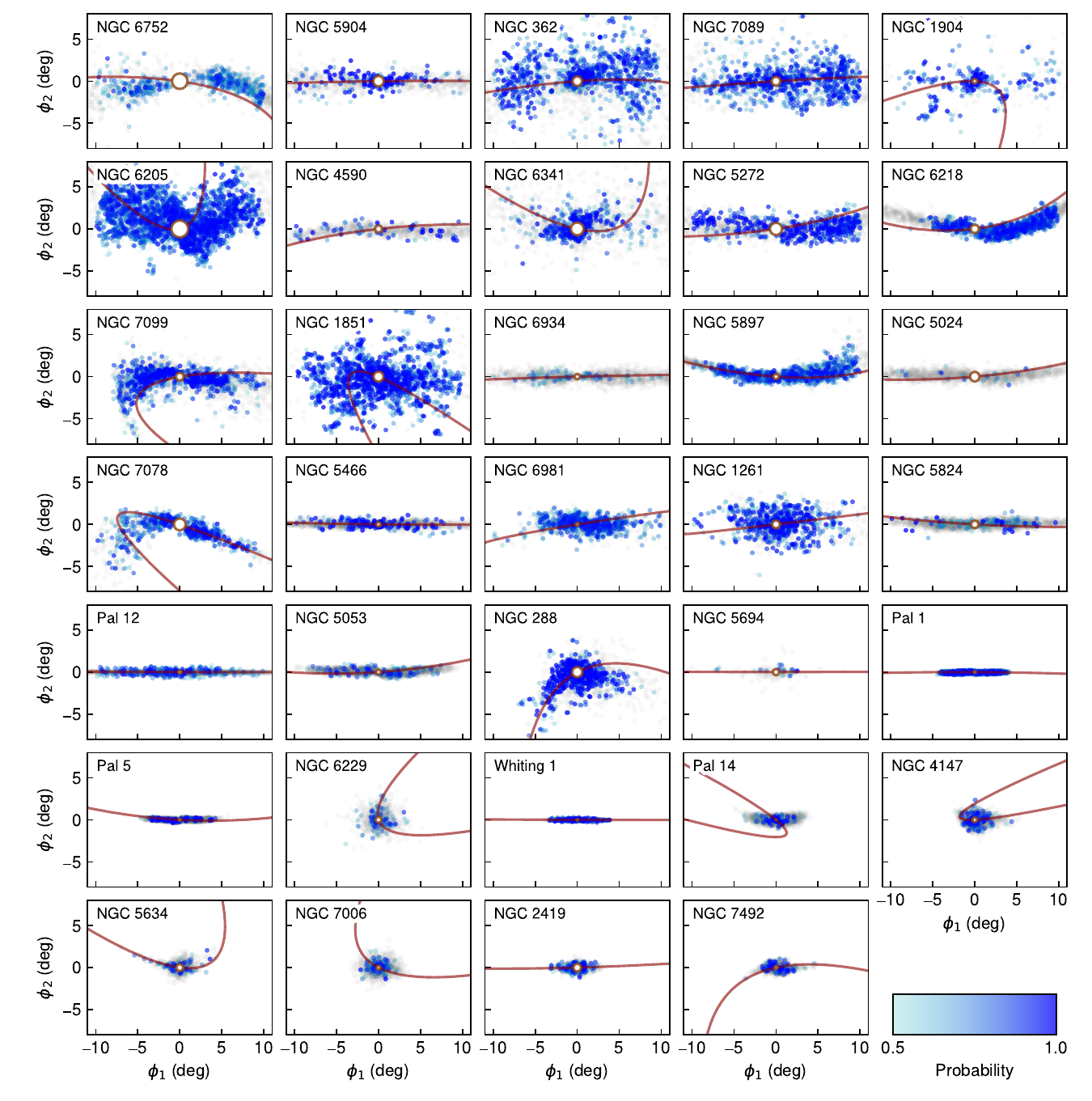}
    \caption{Detections of stream members (blue circles) around 34 MW GCs in the high-quality sample ($A_V<0.6$ and $N_{\rm bg}<6\times10^6$). We show these streams in the great circle frame ($\phi_1$--$\phi_2$) centered on the progenitor GC. Streams are placed in the descending order of the length $r_{90}$. Each star is color-coded by the stream probability, as indicated by the colorbar. The tidal radius of the GC is shown as the brown circle. We show orbits of progenitor GCs as solid brown curves, projected in the same great circle frame. For comparison, we also show the simulated streams (gray symbols).}
    \label{fig:mw_streams}
\end{figure*}

In Table~\ref{tab:properties}, we list the key properties of the 34 streams in our high-quality sample. We provide information about the low-quality sample in Appendix~\ref{sec:low_quality}, Table~\ref{tab:properties_low_quality}. We also plot the high-quality sample in the great circle frame $\phi_1$--$\phi_2$ in Fig.~\ref{fig:mw_streams}. The $\phi_1$--$\phi_2$ frame is defined such that the progenitor GC is at $(0,0)$, and the stream is elongated along $\phi_1$. We achieve this by rotating the sky coordinates to minimize the standard deviation of $\phi_2$ for simulated tracer particles. We rank the streams in Fig.~\ref{fig:mw_streams} by the length of simulated streams, which is characterized by the 90th percentile of angular separation, $r_{90}$. Since we only detect streams within the $10^\circ$ search radius around the GC, there are 17 streams with $r_{90}$ exceeding this radius. In \S\ref{sec:extended_detection}, we perform a followup detection for these streams by extending the search radius to $20^\circ$.

Many streams are wider or more ``irregular'' than the visual expectation that GC streams are thin and long. For example, NGC 4147's stream is almost a circular blob with similar spread in $\phi_1$ and $\phi_2$. However, these streams still have sufficiently distinct distribution in the proper motion space and the color--magnitude space compared to the background stars. They are thus detectable by \textit{StarStream} since we do not make prior assumptions based on the visual expectation of ``regular'' streams.

We also show orbits of progenitor GCs projected onto the great circle frame in Fig.~\ref{fig:mw_streams}. We integrate orbits using the same \texttt{MilkyWayPotential2022} potential. We notice that some streams misalign with the projected orbit of the progenitor GC by $>10^\circ$, such as Palomar~14 (Pal 14). Although the misalignment is expected \citep{sanders_streamorbit_2013,sanders_streamorbit_2013-1,panithanpaisal_breaking_2025} and is likely visually enhanced as these streams have highly eccentric orbits, previous searches of GC streams tended to focus along the GC's motion and preferentially found well-aligned streams such as Pal 5, whose misalignment angle is smaller than $5^\circ$.

The detection of these ``irregular'' or misaligned streams highlights the power of the physics-based modeling of GC streams by \textit{StarStream}. As many GC streams can be dynamically hot or spatially complex depending on the GC’s mass and orbit \citep{amorisco_feathers_2015}, these streams are likely missed by traditional visually-based methods.

\begin{figure*}
    \centering
    \includegraphics[width=\linewidth]{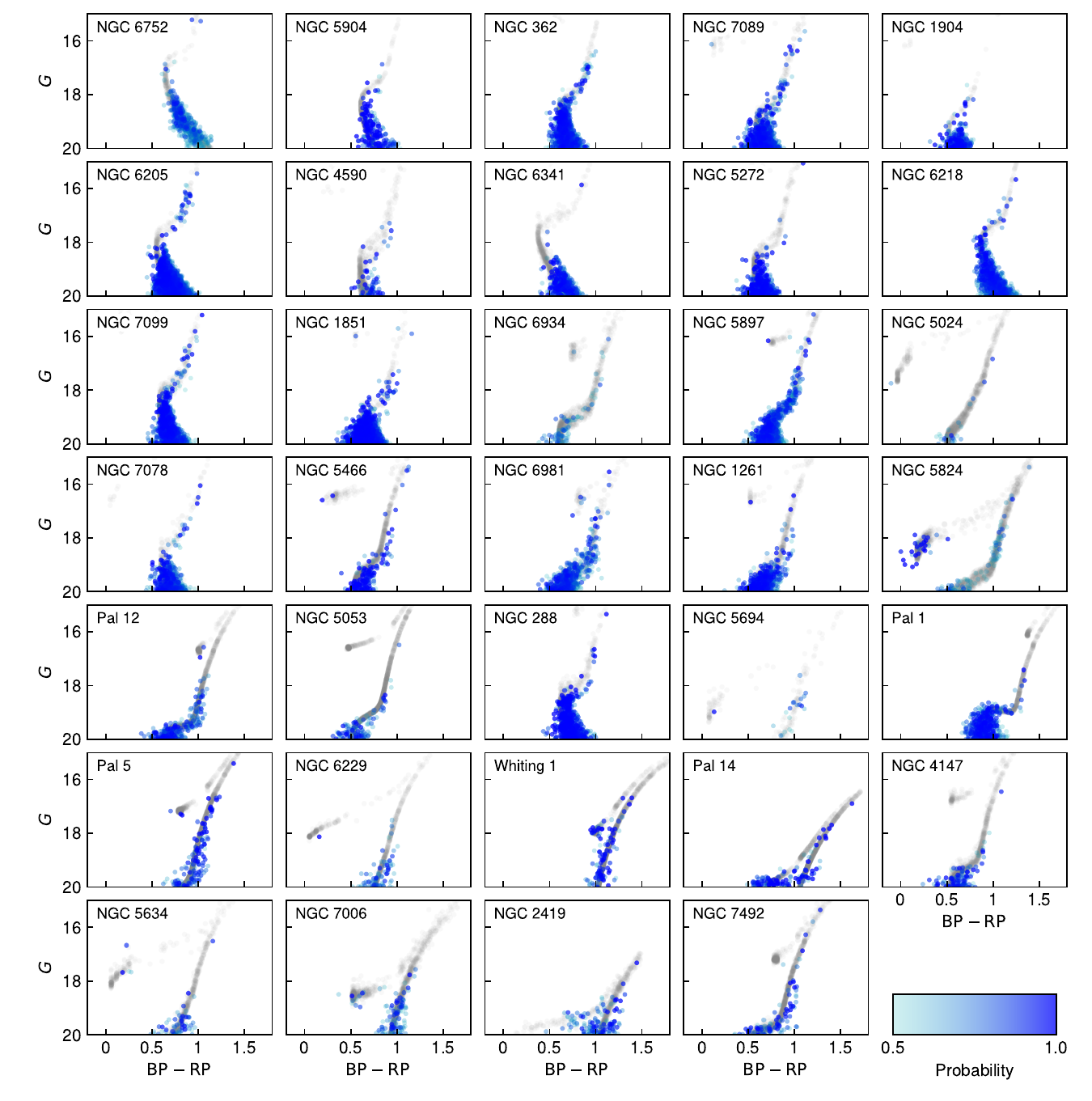}
    \caption{Similar to Fig.~\ref{fig:mw_streams}, but for the color--magnitude space $G$ vs. $\rm BP-RP$.}
    \label{fig:mw_streams_cmd}
\end{figure*}

In Fig.~\ref{fig:mw_streams_cmd}, we show the CMD of detected stream stars using \textit{Gaia} photometry. Most stars are gathered around or below the main sequence turn off. Since \textit{StarStream} takes into account the color and magnitude uncertainties that typically grow with the $G$-magnitude, the spread around isochrone also becomes larger near the fainter end for detected stars. On the other hand, the spread of simulated streams is only due to the distance spread and is thus almost invariant of $G$-magnitude.

\begin{deluxetable*}{lrrrrrrrrrr@{\hskip 8pt}l}
\renewcommand\arraystretch{1.1} 
\tablecaption{Summary of GC and stream properties for the high-quality sample. The Galactic longitude $l$, latitude $b$, heliocentric distance $d_\odot$, cluster mass $M_{\rm GC}$, and 3D half-mass radius $r_{\rm h}$ are taken from the \citet{hilker_galactic_2019} catalog. The extinction $A_V$ is either from D. Massari et al., \citet{harris_catalog_1996} catalog, or \citet{schlegel_maps_1998} map, see \S\ref{sec:obs}. $N_{\rm bg}$ is the number of CMD-selected \textit{Gaia} DR3 stars within the $10^\circ$ search radius. The number of detections $N_{\rm detect}$, tidal frequency $\Omega_{\rm tid}$, and cluster mass loss rate $|\dot{M}|$ are calculated in this work. A plain ASCII version is available at \url{https://github.com/ybillchen/StarStream_DR}.
}
\label{tab:properties}
\tablehead{
\colhead{GC} & \colhead{$l$} & \colhead{$b$} & \colhead{$d_\odot$} & \colhead{$A_V$} & \colhead{$N_{\rm bg}$} & \colhead{$N_{\rm detect}$} &
\colhead{$M_{\rm GC}$} & \colhead{$\Omega_{\rm tid}$} & \colhead{$r_{\rm h}$} & \multicolumn{2}{c}{$|\dot{M}|$} \\
\colhead{} & \colhead{$(^\circ)$} & \colhead{$(^\circ)$} & \colhead{$({\rm kpc})$} & \colhead{} & \colhead{$(10^5)$} & \colhead{} &
\colhead{$(10^5\ {\rm M_\odot})$} & \colhead{$({\rm Gyr^{-1}})$} & \colhead{$({\rm pc})$} & \twocolhead{$({\rm M_\odot\,Myr^{-1}})$}
}
\startdata
NGC 288 & $151.3$ & $-89.4$ & $9.0$ & $0.06$ & $2.2$ & $494$ & $0.962$ & $40.2$ & $8.6$ & $2.7$ & $^{+1.1}_{-0.8}$ \\
NGC 362 & $301.5$ & $-46.2$ & $8.8$ & $0.09$ & $16.6$ & $1030$ & $2.520$ & $42.2$ & $3.4$ & $10.3$ & $^{+11.2}_{-5.4}$ \\
NGC 1261 & $270.5$ & $-52.1$ & $16.4$ & $0.03$ & $3.7$ & $560$ & $1.720$ & $24.5$ & $4.9$ & $12.7$ & $^{+6.1}_{-4.1}$ \\
NGC 1851 & $244.5$ & $-35.0$ & $11.9$ & $0.12$ & $5.8$ & $1483$ & $2.830$ & $25.9$ & $3.2$ & $18.7$ & $^{+12.0}_{-7.3}$ \\
NGC 1904 & $227.2$ & $-29.4$ & $13.1$ & $0.03$ & $7.5$ & $263$ & $1.810$ & $27.6$ & $4.3$ & $19.1$ & $^{+15.1}_{-8.4}$ \\
NGC 2419 & $180.4$ & $25.2$ & $88.5$ & $0.25$ & $8.2$ & $138$ & $7.830$ & $4.4$ & $26.4$ & \multicolumn{2}{c}{$-$} \\
NGC 4147 & $252.8$ & $77.2$ & $18.5$ & $0.06$ & $2.4$ & $108$ & $0.451$ & $20.1$ & $3.5$ & $2.5$ & $^{+1.1}_{-0.8}$ \\
NGC 4590 & $299.6$ & $36.1$ & $10.4$ & $0.16$ & $9.5$ & $68$ & $1.280$ & $13.7$ & $7.3$ & $1.1$ & $^{+1.0}_{-0.5}$ \\
NGC 5024 & $333.0$ & $79.8$ & $18.5$ & $0.06$ & $2.7$ & $25$ & $5.020$ & $18.2$ & $10.0$ & $1.3$ & $^{+0.7}_{-0.5}$ \\
NGC 5053 & $335.7$ & $78.9$ & $17.5$ & $0.03$ & $2.8$ & $113$ & $0.628$ & $20.5$ & $17.0$ & $3.4$ & $^{+1.5}_{-1.1}$ \\
NGC 5272 & $42.2$ & $78.7$ & $10.2$ & $0.03$ & $2.7$ & $503$ & $4.090$ & $27.6$ & $5.5$ & $6.2$ & $^{+2.7}_{-1.9}$ \\
NGC 5466 & $42.1$ & $73.6$ & $16.1$ & $0.00$ & $2.9$ & $162$ & $0.561$ & $8.6$ & $13.8$ & $4.1$ & $^{+1.8}_{-1.3}$ \\
NGC 5634 & $342.2$ & $49.3$ & $26.0$ & $0.16$ & $7.1$ & $62$ & $2.470$ & $22.5$ & $7.7$ & $10.8$ & $^{+8.7}_{-4.8}$ \\
NGC 5694 & $331.1$ & $30.4$ & $34.8$ & $0.28$ & $22.3$ & $20$ & $2.690$ & $7.6$ & $4.3$ & $7.3$ & $^{+10.1}_{-4.2}$ \\
NGC 5824 & $332.6$ & $22.1$ & $31.7$ & $0.40$ & $49.2$ & $131$ & $7.460$ & $10.8$ & $6.3$ & $62.5$ & $^{+116.0}_{-40.6}$ \\
NGC 5897 & $342.9$ & $30.3$ & $12.6$ & $0.37$ & $21.2$ & $764$ & $1.670$ & $49.6$ & $10.9$ & $16.9$ & $^{+20.5}_{-9.3}$ \\
NGC 5904 & $3.9$ & $46.8$ & $7.5$ & $0.09$ & $7.1$ & $176$ & $3.920$ & $17.7$ & $5.6$ & $2.4$ & $^{+1.7}_{-1.0}$ \\
NGC 6205 & $59.0$ & $40.9$ & $7.4$ & $0.03$ & $7.1$ & $2209$ & $4.840$ & $56.4$ & $5.2$ & $16.7$ & $^{+12.0}_{-7.0}$ \\
NGC 6218 & $15.7$ & $26.3$ & $5.1$ & $0.59$ & $23.7$ & $806$ & $1.060$ & $86.2$ & $4.1$ & $3.9$ & $^{+5.0}_{-2.2}$ \\
NGC 6229 & $73.6$ & $40.3$ & $30.1$ & $0.03$ & $6.5$ & $79$ & $2.470$ & $17.1$ & $4.8$ & $25.1$ & $^{+24.8}_{-12.5}$ \\
NGC 6341 & $68.3$ & $34.9$ & $8.5$ & $0.03$ & $7.3$ & $356$ & $2.730$ & $48.2$ & $3.6$ & $3.1$ & $^{+2.3}_{-1.3}$ \\
NGC 6752 & $336.5$ & $-25.6$ & $4.1$ & $0.12$ & $41.1$ & $478$ & $2.610$ & $71.2$ & $4.8$ & $2.9$ & $^{+4.6}_{-1.8}$ \\
NGC 6934 & $52.1$ & $-18.9$ & $15.7$ & $0.31$ & $52.9$ & $83$ & $1.500$ & $9.4$ & $4.7$ & $3.9$ & $^{+6.8}_{-2.5}$ \\
NGC 6981 & $35.2$ & $-32.7$ & $16.7$ & $0.16$ & $17.8$ & $628$ & $0.812$ & $22.1$ & $5.8$ & $18.0$ & $^{+20.3}_{-9.5}$ \\
NGC 7006 & $63.8$ & $-19.4$ & $39.3$ & $0.16$ & $42.6$ & $108$ & $1.320$ & $9.0$ & $6.6$ & $81.5$ & $^{+132.3}_{-50.4}$ \\
NGC 7078 & $65.0$ & $-27.3$ & $10.7$ & $0.31$ & $15.5$ & $504$ & $5.180$ & $41.7$ & $3.7$ & $5.9$ & $^{+6.3}_{-3.0}$ \\
NGC 7089 & $53.4$ & $-35.8$ & $11.7$ & $0.16$ & $10.1$ & $814$ & $6.240$ & $27.1$ & $4.8$ & $14.6$ & $^{+12.7}_{-6.8}$ \\
NGC 7099 & $27.2$ & $-46.8$ & $8.5$ & $0.12$ & $7.0$ & $728$ & $1.210$ & $55.5$ & $4.3$ & $4.3$ & $^{+3.1}_{-1.8}$ \\
NGC 7492 & $53.4$ & $-63.5$ & $24.4$ & $0.00$ & $3.4$ & $96$ & $0.197$ & $19.0$ & $10.6$ & $9.0$ & $^{+5.3}_{-3.3}$ \\
Pal 1 & $130.1$ & $19.0$ & $11.2$ & $0.46$ & $17.0$ & $445$ & $0.009$ & $17.1$ & $3.4$ & $4.9$ & $^{+5.5}_{-2.6}$ \\
Pal 5 & $0.8$ & $45.9$ & $21.9$ & $0.09$ & $8.5$ & $131$ & $0.134$ & $21.5$ & $27.6$ & $10.1$ & $^{+8.3}_{-4.6}$ \\
Pal 12 & $30.5$ & $-47.7$ & $18.5$ & $0.06$ & $7.3$ & $228$ & $0.062$ & $6.7$ & $10.5$ & $8.7$ & $^{+6.4}_{-3.7}$ \\
Pal 14 & $28.7$ & $42.2$ & $73.6$ & $0.12$ & $7.9$ & $117$ & $0.191$ & $4.0$ & $37.7$ & \multicolumn{2}{c}{$-$} \\
Whiting 1 & $161.6$ & $-60.6$ & $30.6$ & $0.09$ & $2.1$ & $99$ & $0.014$ & $5.4$ & $16.5$ & $26.0$ & $^{+21.9}_{-11.9}$
\vspace{2mm}
\enddata
\end{deluxetable*}

\subsection{Extended detection for long streams}
\label{sec:extended_detection}

We find 17 streams in the high-quality sample that extend beyond the default search radius, or $r_{\rm 90}>10^\circ$. To obtain a more complete detection for these streams, we rerun \textit{StarStream} for these streams in the $10^\circ-20^\circ$ annulus around each GC. For consistency, we keep all other parameters of the method the same as those inside $10^\circ$. 

\begin{figure}
    \centering
    \includegraphics[width=\linewidth]{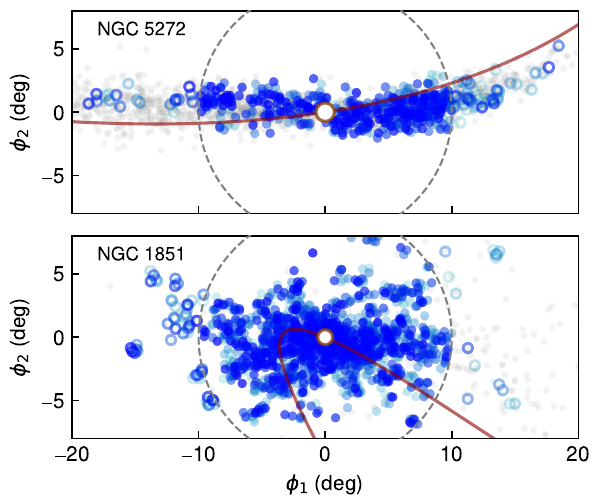}
    \caption{Similar to Fig.~\ref{fig:mw_streams}, but for extended streams with $r_{90}>10^\circ$. Only streams with more than 10 extended detections outside the original $10^\circ$ search radius are shown here, with the extended detections shown as open symbols. The original search radius is marked as dashed circles in each panel.}
    \label{fig:mw_streams_long}
\end{figure}

In Fig.~\ref{fig:mw_streams_long}, we show the two streams, NGC 5272 and 1851, with more than 10 new detections outside $10^\circ$. Note that the number of background stars approximately triples in the annulus, significantly reducing the signal-to-noise ratio in this region. Therefore, we should not expect the completeness and purity to stay unchanged in the extended area.

\subsection{Comparison with previous detections}
\label{sec:comparison}

\citet{ibata_charting_2024} applied the \texttt{STREAMFINDER} algorithm to \textit{Gaia} DR3 and provided the most complete catalog of GC streams prior to this work. They found 16 GCs that are associated with stellar streams. However, three of their streams, NGC 4590, 5139, and 5904, are not directly connected to the GC and thus cannot be compared with our results. For the remaining 13 streams, 8 are in our high-quality sample: NGC 288, 1261, 1851, 5466, 6341, 7089, 7099, and Pal 5, whereas 5 are in our low-quality sample: NGC 2298, 2808, 3201, 6101, and 6397. For the high-quality sample, our method always yields higher $N_{\rm detect}$ within the $10^\circ$ radius by $20-6000\%$, with a median of $300\%$. For the low-quality sample, our method detects more stars for NGC 2298, 2808, 6101, and fewer for the remaining two. However, since the low-quality sample has large background contamination and extinction, either method requires follow-up observations for further confirmation.

For Pal 5 specifically, we detect 131 member stars while \texttt{STREAMFINDER} detects 109. However, since our method focuses on the stream segment released in the last 1~Gyr, our detections are all concentrated in the inner $5^\circ$. For this region, \citet{ibata_charting_2024} reported 76 stars. The 70\% improvement of our method is remarkable since Pal 5 is one of the most complete streams known to date.

In addition to automated stream searches, there has also been detections of streams or tidal features around individual clusters, including five GCs by \citet{sollima_eye_2019}, four GCs by \citet{shipp_stellar_2018}, NGC 288 \citep{piatti_extended_2018,kaderali_rediscovering_2019,grillmair_multiple_2025}, NGC 1261 and 1904 \citep{awad_s5_2025}, NGC 4590 \citep{palau_tidal_2021}, NGC 5024 and NGC 5053 \citep{yuan_low-mass_2020}, NGC 5272 \citep{yang_spectacular_2023}, NGC 5466 \citep{jensen_uncovering_2021}, NGC 5904 \citep{grillmair_detection_2019}, NGC 6341 \citep{thomas_hidden_2020}, and NGC 7089 \citep{grillmair_extended_2022}. We do not present a detailed, stream-by-stream comparison here, as most of these detections are already in the \citet{ibata_charting_2024} catalog. Moreover, many of these studies do not release data products that are directly comparable with our data, and they typically emphasize purity over completeness through strict selection criteria. However, we still find that the stream morphologies and kinematics reported in these works are consistent with our detections within the $10^\circ$ search radius.

Nevertheless, some individually detected streams are not in \citet{ibata_charting_2024}. For example, \citet{awad_s5_2025} reported streams around NGC 1261 and NGC 1904 using the Southern Stellar Stream Spectroscopic Survey \citep[S$^5$;][]{li_southern_2019}, and NGC 1904 is not in \citet{ibata_charting_2024}. By examining stream stars outside the tidal radius, we find that 53\% (NGC~1261) and 27\% (NGC~1904) of their reported stream stars are recovered in our catalog. Given that the completeness of our method is $\sim$60\% for NGC 1261 and $\sim$30\% for NGC 1904 (as indicated from Fig.~\ref{fig:mock_real}), we can infer a purity $\sim$90\% for their detections. On the other hand, our completeness is likely more than three times higher, as we detect an order of magnitude more stream stars around these clusters. This level of agreement is encouraging, especially given that we do not incorporate radial velocities or metallicity information from spectroscopic surveys.

Recently, \citet{kuzma_constructing_2025} identified potential tidal structures within $5^\circ$ around 22 MW GCs using the Pristine-\textit{Gaia} synthetic catalog from Pristine Data Release 1 \citep{starkenburg_pristine_2017,martin_pristine_2024}. 12 of them overlap our high-quality sample: NGC 362, 1261, 1851, 1904, 5272, 6205, 6341, 6934, 6981, 7078, 7089, 7099. It is worth noting that many of their tidal structures do not show stream-like feature. This agrees with our finding that although many GC streams are ``irregular'', the tidal streams around GCs are more common than was previously detected.

\section{Mass loss rate of globular clusters}
\label{sec:mass_loss}

\subsection{Calculation of mass loss rate}

Using the stream stars identified by \textit{StarStream}, we next measure the orbit-averaged mass loss rate $\dot{M}$ of the progenitor GCs. We use an updated approach from \citet{chen_stellar_2025} starting from their Eq.~(3):
\begin{equation}
    \dot{M}\approx\frac{\dot{N}_{\rm tracer}\, M_{\rm sel}}{\iint f_{\rm sel}(\phi_{1},\phi_{2})\, n_{\rm tracer}(\phi_{1},\phi_{2})\, d\phi_1 d\phi_2}
    \label{eq:mass_loss_exact}
\end{equation}
where $M_{\rm sel}$ is the total detected mass of stream stars. $M_{\rm sel}$ is equivalent to the total mass selected by the effective spatial selection function $f_{\rm sel}(\phi_1,\phi_2)$ of \textit{StarStream}. $\dot{N}_{\rm tracer}$ is the number ejection rate of tracer particles above the detection limit. The integral is over the number density of tracer particles $n_{\rm tracer}(\phi_{1},\phi_{2})$ weighted by the selection function. 

\citet{chen_stellar_2025} approximates $f_{\rm sel}$ as a Gaussian tube around the stream track when analyzing the \citet{ibata_charting_2024} stream catalog. Here, however, we can make further simplification considering the quantified detection ratio $f_{\rm detect}$ between the expectation of the number of detections $N_{\rm detect}$ by \textit{StarStream} and the true number $N_{\rm true}$ \citepalias[see][]{chen_starstream_2025-1}. In Appendix~\ref{sec:unbiased_mass_loss_rate}, we prove that simply replacing the integral in Eq.~(\ref{eq:mass_loss_exact}) by the total number of tracer particles $N_{\rm tracer}$ times $f_{\rm detect}$ yields an unbiased estimate of the mass loss rate:
\begin{equation}
    \dot{M}\approx f_{\rm detect}^{-1}\frac{\dot{N}_{\rm tracer}\, M_{\rm sel}}{N_{\rm tracer}}
    \label{eq:mass_loss}
\end{equation}
where $f_{\rm detect}\equiv N_{\rm detect}/N_{\rm true}$ varies with individual streams. Using the tests by \citetalias{chen_starstream_2025-1}, we find that $f_{\rm detect}$ is always centered at 0.9 for the high-quality sample, with a log-normal scatter that depends on the background density $N_{\rm bg}$. In Appendix~\ref{sec:unbiased_mass_loss_rate}, we approximate the standard deviation of the log-normal scatter $\sigma_{\log f}$ as a linear function of $\log_{10}N_{\rm bg}$: $\sigma_{\log f}$ increases from 0.15~dex to 0.5~dex when $\log_{10}N_{\rm bg}$ increases from 5.5 to 7, and is fixed to $\sigma_{\log f}=0.15$~dex below $\log_{10}N_{\rm bg}<5.5$ (see Fig.~\ref{fig:fdetect_vs_nbg}). Since the purity and completeness of the low-quality sample are both low with large variation, the scatter can be more than 1~dex. Therefore, we exclude the low-quality sample for the calculation of the mass loss.

Note that $M_{\rm sel}$ and $N_{\rm tracer}$ only account for tracer particles within the $10^\circ$ search radius. The detection ratio is not calibrated outside $10^\circ$ and is likely much lower. We thus exclude extended stream segments outside this radius (see \S\ref{sec:extended_detection}) for the subsequent calculation.

For real observations, $M_{\rm sel}$ is not simply the sum of the masses of individual stars, as it must account for stars below the detection limit. Following \citet{chen_stellar_2025}, we introduce a correction factor $w_i$ for the $i$-th star to account for the missing stellar mass
\begin{equation}
    M_{\rm sel}\approx\sum_{i=1}^{N_{\rm obs}} m_i \, w_i
    \label{eq:estimate}
\end{equation}
where
\begin{equation*}
    w_i=w(m_{{\rm min},i})\equiv\frac{\int_{m_{\rm limit}}^{m_{\rm max}} m\, \psi(m)\, dm}{\int_{m_{{\rm min},i}}^{m_{\rm max}}m\, \psi(m)\, dm}
\end{equation*}
in which the upper integral limit $m_{\rm max}$ is the maximum mass of surviving stars at the current age. We set the lower integration limit of the numerator to the hydrogen-burning threshold $m_{\rm limit} = 0.08\ M_\odot$, and the lower integration limit of the denominator to the minimum detectable stellar mass at the heliocentric distance of this star $m_{{\rm min},i}\equiv m_{\rm min}(d_{\odot,i})$. The stellar mass function of the stream is denoted by $\psi(m)$. Following \citet{chen_stellar_2025}, we assume $\psi(m)$ follows the same power-law function as the progenitor GC, with the slope measured by \citet{baumgardt_evidence_2023}. These authors also attempted to fit the mass function with broken power-law functions. However, only a subset of our sample GCs have measured slopes for the broken power-law. For these GCs, the two forms of mass functions deviate only  by $<20\%$ when calculating the correction factor.

Since we focus on the $10^\circ$ cone around each GC, the heliocentric distance of each stream member $d_{\odot,i}$ is very close to the distance to the center GC, $d_{\odot,\rm GC}$. Using the simulated stream, we verify that the standard deviation of distances is typically $0.025$~dex (at most $0.08$~dex). Therefore, the distance spread contributes negligibly to the total uncertainty since the intrinsic scatter due to the detection method itself is already $>0.15$~dex (see Appendix~\ref{sec:unbiased_mass_loss_rate}). It is thus reasonable to adopt $d_{\odot,i} \approx d_{\odot,\rm GC}$. Correspondingly, we can define $m_{{\rm min},i} = m_{\rm min}(d_{\odot,\rm GC})\equiv  m_{\rm min,GC}$ and $w_i=w(m_{\rm min,GC})\equiv w_{\rm GC}$. Eq.~(\ref{eq:estimate}) is thus simplified to
\begin{equation*}
    M_{\rm sel}\approx\sum_{i=1}^{N_{\rm obs}} m_i \, w_{\rm GC} = \left(\sum_{i=1}^{N_{\rm obs}} m_i\right)w_{\rm GC} \equiv M_{\rm obs}\, w_{\rm GC}.
\end{equation*}
The final formula for mass loss rate is given by
\begin{equation}
    \dot{M}\approx f_{\rm detect}^{-1}\frac{\dot{N}_{\rm tracer}\, M_{\rm obs}\, w_{\rm GC}}{N_{\rm tracer}}.
    \label{eq:mass_loss_final}
\end{equation}

It should be noted that the correction factor $w_{\rm GC}$ becomes extremely large ($>1000$) and is sensitive to the parameters for the isochrone model when the GC is far away from the observer. We find that varying the age of the isochrone from $7-13$~Gyr alters $w_{\rm GC}$ by several orders of magnitudes when $d_{\odot,\rm GC}\gtrsim 60$~kpc, where the main sequence turnoff is far below \textit{Gaia}'s detection limit. Therefore, we exclude NGC 2419 and Pal 14 since they both have $d_{\odot,\rm GC}>60$~kpc. Their inferred mass loss rates likely have too large uncertainties. Except for these two, we verify that the scatter of detection ratio $\sigma_{\log f}$ in Eq.~(\ref{eq:mass_loss}) is dominant over all other potential sources of uncertainties, including the uncertainty in $w_{\rm GC}$ and the Poisson's error for small numbers. However, we still include these sources of uncertainties following \citet{chen_stellar_2025} for the subsequent analysis.

We also note that several GCs in our catalog are associated with the Sagittarius (Sgr) stream \citep{bellazzini_globular_2020,vasiliev_tango_2021}, including NGC 6715 (the nucleus), Arp 2, Ter 7, Ter 8, Whiting 1, and Pal 12. Among these systems, Whiting 1 and Pal 12 are in our high-quality sample analyzed here. Our selection may include stars from the Sgr stream, whose spatial, proper motion, color, and magnitude distributions are similar to those of these GCs. As a result, the inferred number of stream stars may be overestimated, and we therefore interpret our derived $\dot{M}$ as upper limits for these clusters.

\subsection{Mass loss rate vs. other cluster properties}

\begin{figure*}
    \centering
    \includegraphics[width=\linewidth]{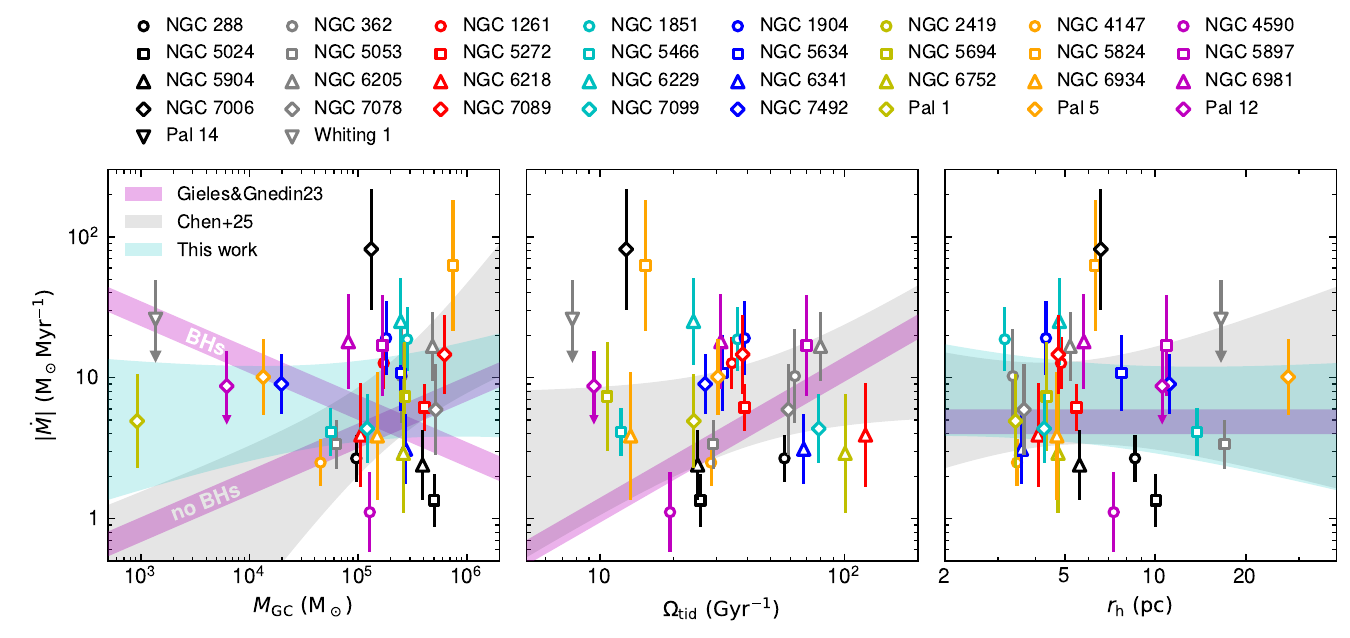}
    \caption{Mass loss rates of 34 streams in the high-quality sample, plotted against $M_{\rm GC}$ (\textit{left}), $\Omega_{\rm tid}$ (\textit{middle}), and $r_{\rm h}$ (\textit{right}), with uncertainties shown as errorbars. The best-fit relation for these measurements are shown as light blue shaded regions. For comparison, we also show the BHs and no BHs models from \citet[][with $(a,b,c)=(\pm1/3,1,0)$ and $|\dot{M}_{\rm ref}|=30-45\ {\rm M_\odot\,Myr^{-1}}$]{gieles_mass-loss_2023} and the best-fit relations in \citet{chen_stellar_2025} as magenta and gray shaded regions, respectively. Note that $\Omega_{\rm tid}$ in \citet{chen_stellar_2025} is smaller by a constant $\sqrt{2}$, which we have accounted for in the comparison here.}
    \label{fig:mass_loss}
\end{figure*}

Figure~\ref{fig:mass_loss} shows the measured mass loss rate for the high-quality sample as functions of GC's mass $M_{\rm GC}$, effective tidal frequency $\Omega_{\rm tid}$, and 3D half-mass radius $r_{\rm h}$. The values of $M_{\rm GC}$ and $r_{\rm h}$ are directly taken from the \citet{hilker_galactic_2019} catalog, while $\Omega_{\rm tid}$ is approximated by $\sqrt{2}$ times the orbital frequency, which is characterized by the time between two adjacent apocentric passages. This definition is consistent with \citet{gieles_mass-loss_2023}, who used the singular isothermal sphere profile and adopted \citet{renaud_evolution_2011}'s definition based on eigenvalues of the tidal tensor. The orbital frequency is computed via integrating the GC's orbit in our Galactic potential model. These values are listed in Table~\ref{tab:properties}.

We find that most GCs have $\dot{M}=1-100\ {\rm \Msun\,Myr^{-1}}$. We do not observe any strong correlation between $\dot{M}$ and other properties of the GC. It should be noted that we do not include streams with fewer than 10 detections in the sample, which excludes streams with very low mass loss rates. For the $\dot{M}$--$M_{\rm GC}$ relation and the $\dot{M}$--$r_{\rm h}$ relation, the exclusion of these streams biases the entire relation upward. 

However, the $\dot{M}$--$\Omega_{\rm tid}$ relation is affected more because the low-$\Omega_{\rm tid}$ GCs tend to reside at large radii. These GCs are more likely to be excluded by the same $N_{\rm detect}>10$ criterion unless their mass loss rates are proportionally higher. Therefore, the slope of the $\dot{M}$--$\Omega_{\rm tid}$ relation is likely biased low. For this reason, we do not analyze the $\dot{M}$--$\Omega_{\rm tid}$ relation for the remainder of the work. To quantitatively study the correlation between $\dot{M}$ and the other two properties, we fit $\dot{M}$ as a multivariate power-law function,
\begin{equation}
    \dot{M}_{\rm fit} = \dot{M}_{\rm ref}\left(\frac{M_{\rm GC}}{10^5\Msun}\right)^a\left(\frac{r_{\rm h}}{5\ {\rm pc}}\right)^c
    \label{eq:powerlaw}
\end{equation}
where we choose different anchor points from \citet{chen_stellar_2025} to better describe the average of each quantity\footnote{We skip the symbol $b$ for consistency with \citet{chen_stellar_2025}, where $b$ is the slope for the $\dot{M}$--$\Omega_{\rm tid}$ relation.}. Using maximum likelihood estimation allowing for intrinsic scatter, we obtain the best-fit parameters:
\begin{equation}
    \begin{array}{rl}
        |\dot{M}_{\rm ref}| &= 6.8^{+1.6}_{-1.3}\Msun\,{\rm Myr^{-1}} \\
        \sigma_{\rm int} &= 0.27\pm0.06\ {\rm dex} \\
        a &= 0.09\pm0.18 \\
        c &= -0.20\pm0.39
    \end{array}
    \label{eq:bestfit}
\end{equation}
where the uncertainties are obtained from 1000 bootstrap resampling. We find that the $\dot{M}$--$M_{\rm GC}$ slope $a$ is consistent with zero within the standard deviation, and the value lies between the no black holes (BHs) model ($a=1/3$) and BHs models ($a=-1/3$, assuming the initial mass $M_{\rm i}=2\times10^5\Msun$) of \citet{gieles_mass-loss_2023}. Our $a$ is lower than that in \citet[][$a=0.66\pm 0.37$]{chen_stellar_2025} by $1.6\sigma$. The $\dot{M}$--$r_{\rm h}$ slope is slightly below but still consistent with zero. It is also consistent with \citet[][$c=0.12\pm 0.72$]{chen_stellar_2025}. We also observe larger intrinsic scatter compared to \citet{chen_stellar_2025}. This is likely because our measurement includes the uncertainty of the detection method itself, whereas \citet{chen_stellar_2025} assumed zero uncertainty for the detection method. Although trends can differ, the amplitude of $\dot{M}$ in this work is consistent with \citet{gieles_mass-loss_2023} and \citet{chen_stellar_2025} within the intrinsic scatter across the full ranges of $M_{\rm GC}$, $\Omega_{\rm tid}$, and $r_{\rm h}$. 

In particular, we measured $|\dot{M}|\approx10\ {\rm M_\odot\,Myr^{-1}}$ for Pal 5. This is higher than that in \citet[][$|\dot{M}|\approx3\ {\rm M_\odot\,Myr^{-1}}$]{chen_stellar_2025} partially because of the 70\% increase of $N_{\rm detect}$ (as mentioned in \S\ref{sec:comparison}). However, this increase is insufficient to explain the 230\% increase of $\dot{M}$. Since \citet{chen_stellar_2025} measured the average mass loss rate over the past $\sim6$~Gyr for Pal 5 while this work only measures the past 1~Gyr, the different values likely suggest an accelerating mass loss rate for Pal 5. This scenario is consistent with the prediction by \citet{gieles_supra-massive_2021}, who showed that Pal 5's potential high BH abundance ($f_{\rm BH}\approx 20\%$) could accelerate the mass loss rate from the initial $5-10\ {\rm M_\odot\,Myr^{-1}}$ to $10-20\ {\rm M_\odot\,Myr^{-1}}$ near the end of its lifetime.

We also notice several low-mass but high-$\dot{M}$ GCs similar to Pal 5, including NGC 7492, Pal 12, and Whiting~1. These GCs have $M_{\rm GC}\lesssim2\times10^4\Msun$ but half-mass radii above average, $r_{\rm h}\gtrsim10$~pc, making them much ``fluffier'' than average GCs. Their high mass loss rates place them in closer agreement with \citet{gieles_mass-loss_2023}'s BHs models ($a=-1/3$), while they are seemingly outlier of other models. \citet{gieles_supra-massive_2021} suggested that these GCs are also BH-rich and are reaching complete tidal dissolution. They also predicted higher-than-average mass loss rates for these GCs, $|\dot{M}|\approx 10\ {\rm M_\odot\,Myr^{-1}}$. Our measurements agree with their prediction, providing observational support for the BH-rich scenario.

\section{Summary and discussion}
\label{sec:summary}

We report 87 GC streams detected by the \textit{StarStream} method in \textit{Gaia} DR3. Our catalog includes a high-quality sample of 34 streams with $A_V<0.6$ and $N_{\rm bg}<6\times10^6$ within the $10^\circ$ search radius (Figs.~\ref{fig:mw_streams}, \ref{fig:mw_streams_cmd}, \ref{fig:mw_streams_long} and Table~\ref{tab:properties}), and a low-quality sample of 53 streams with higher extinction or background density (Table~\ref{tab:properties_low_quality}). Based on our validation tests on a similar mock dataset (Fig.~\ref{fig:mock_real}), our selection criteria for the high-quality sample lead to both median completeness and purity above 50\%. Given these metrics, we provide the most complete catalog of GC streams with quantified detection quality.

This discovery significantly improves our knowledge of GC streams, as even just the high-quality sample doubles the number of known GC streams to date. Moreover, this sample includes around 75\% of the total of 44 GCs with the same selection criteria, which approximately corresponds to the high-latitude region $|b|\gtrsim20^\circ$. For the remaining 25\%, we have too few detections ($N_{\rm detect}\leq10$) to confirm neither their existence nor absence. This remarkable recovery fraction suggests that tidal streams are common around GCs. However, some streams can have ``irregular'' morphology or deviate the progenitor's orbit, contradicting the visual expectation that streams are thin features elongated along the progenitor's orbit. Therefore, the physics-based modeling of GC streams is necessary to uncover these streams.

Our validation tests verified the near-unity detection ratio for the high-quality sample, enabling us to obtain first-ever unbiased estimate of the stream density. Based on this density, we calculate the mass loss rate of their progenitor GCs (Fig.~\ref{fig:mass_loss}). However, the detection ratio has a log-normal scatter varying between 0.15~dex and 0.5~dex depending on the background density (Appendix~\ref{sec:unbiased_mass_loss_rate} and Fig.~\ref{fig:fdetect_vs_nbg}), dominating the uncertainty of $\dot{M}$. As discussed in \citetalias{chen_starstream_2025-1}, next generation wide-field surveys are likely to reduce this scatter by providing either additional independent observables (such as metallicity and line-of-sight velocity) or lower photometry and proper motion uncertainties.

By fitting $\dot{M}$ as a multivariate power-law function of the GC mass $M_{\rm GC}$ and half-mass radius $r_{\rm h}$, we observe the slopes for both quantities are consistent with zero with large scatters. We also notice several ``fluffy'' GCs with low mass and large radius. These GCs show high $|\dot{M}|\approx10\ {\rm M_\odot\,Myr^{-1}}$. They are consistent with the BH-rich scenario of \citet{gieles_supra-massive_2021}, where the BH population enhances both the half-mass radius and mass loss rate near the end of the GC's lifetime.

To facilitate access by the broader community, we publicly release our detection results on GitHub at \url{https://github.com/ybillchen/StarStream_DR}, accompanied with an example notebook for basic instructions.

\section*{Acknowledgments}
We thank Monica Valluri, Eric Bell, and Colin Holm-Hansen for insightful discussions. We thank Davide Massari for sharing updated extinction measurements for select GCs. 
We also thank the reviewer for suggestions and comments, which have improved the quality of this work.
YC and OYG were supported in part by National Aeronautics and Space Administration through contract NAS5-26555 for Space Telescope Science Institute programs HST-AR-16614 and JWST-GO-03433.
This research benefited from the Gravity in the Local Group conference hosted by the McWilliam's Center for Cosmology and Astrophysics, Carnegie Mellon University.

\software{
\textit{StarStream} \citep{chen_starstream_2025-1}, 
\texttt{agama} \citep{vasiliev_agama_2019}, 
\texttt{numpy} \citep{harris_array_2020}, 
\texttt{matplotlib} \citep{hunter_matplotlib_2007}, 
\texttt{scipy} \citep{virtanen_scipy_2020}, 
\texttt{astropy} \citep{the_astropy_collaboration_astropy_2018}, 
\texttt{gala} \citep{price-whelan_gala_2017,price-whelan_adrngala_2024}, 
\texttt{pandas} \citep{the_pandas_development_team_pandas-devpandas_2024},
\texttt{dustmaps} \citep{green_dustmaps_2018}
}

\appendix
\vspace{-6mm}

\section{Special treatment for NGC 5024 and 5053}
\label{sec:NGC_5024_5053}

The method outlined in \S\ref{sec:StarStream} assumes that there is no significant overlap between multiple streams. This assumption is not valid for the GC pair NGC 5024 and 5053, which reside near the Galactic pole and are separated by only $\sim1^\circ$. \citet{yuan_low-mass_2020} suggested that the streams associated with these two clusters overlap substantially, although their radial velocities differ by $\sim100\ {\rm km\,s^{-1}}$. To account for this complexity, we adopt a three-component mixture model to describe the probability density function with two distinct streams:
\begin{equation*}
    p({\bm x}) = f_{\rm s1}\, p_{\rm s1}({\bm x}) + f_{\rm s2}\, p_{\rm s2}({\bm x}) + (1-f_{\rm s1}-f_{\rm s2})\, p_{\rm bg}({\bm x}),
\end{equation*}
where indices 1 and 2 correspond to the streams of NGC 5024 and 5053, respectively. The resulting model therefore introduces two free parameters, $f_{\rm s1}$ and $f_{\rm s2}$, which we fit jointly by maximizing the corresponding likelihood function.

Once we obtain the best-fit values of $f_{\rm s1}$ and $f_{\rm s2}$, we compute the probability that star $i$ belongs to stream 1 as
\begin{equation*}
    P_{{\rm s1},i} \equiv 
    \frac{f_{\rm s1}\, p_{\rm s1}({\bm x}_i)}
    {f_{\rm s1}\, p_{\rm s1}({\bm x}_i) + f_{\rm s2}\, p_{\rm s2}({\bm x}_i) + (1-f_{\rm s1}-f_{\rm s2})\, p_{\rm bg}({\bm x}_i)} .
\end{equation*}
The probability $P_{{\rm s2},i}$ is computed in a similar manner. We classify stars with $P_{{\rm s1},i}+P_{{\rm s2},i}>P_{\rm th}$ as stream members. Among these stars, those with $P_{{\rm s1},i}>P_{{\rm s2},i}$ are assigned to stream 1, while the remainder are assigned to stream 2.

As described in \S\ref{sec:method_validation}, our search is restricted to stars within a $10^\circ$ radius of each GC. For the GC pair here, the simulated stream of NGC 5053 is shorter than $10^\circ$ and is therefore fully enclosed within the search radius centered on NGC 5024. Consequently, we search for both streams within $10^\circ$ around NGC~5053. In addition, because either GC passes through the stream of each other, we exclude stars within the tidal radii of both clusters when performing the joint fit.

We find that the likelihood function peaks at $f_{\rm s1}=2.29\times10^{-4}$ (NGC 5024) and $f_{\rm s2}=4.57\times10^{-4}$ (NGC 5053), with no significant degeneracy between the two parameters. This indicates that the presence of both streams is necessary to describe the overall stellar distribution in the surrounding region. This appears to be in tension with \citet{yuan_low-mass_2020}, who classified several stars as belonging to a single, wide stream associated with both GCs. However, their stars are predominantly far from the two clusters ($>10^\circ$), where the individual streams may already be dynamically mixed.

\section{Proof of the unbiased estimate of mass loss rate}
\label{sec:unbiased_mass_loss_rate}

Given the true number density of stream stars $n_{\rm true}(\phi_{1},\phi_{2})$, the true number of stars $N_{\rm true}$ is
\begin{equation*}
    N_{\rm true} = \iint n_{\rm true}\, (\phi_{1},\phi_{2})d\phi_1 d\phi_2.
\end{equation*}
Similarly, given the number density of simulated tracer particles $n_{\rm tracer}(\phi_{1},\phi_{2})$, the total number of tracers $N_{\rm tracer}$ is
\begin{equation*}
    N_{\rm tracer} = \iint n_{\rm tracer}(\phi_{1},\phi_{2})\, d\phi_1 d\phi_2.
\end{equation*}
If we assume that the simulated stream is an unbiased estimate of the true stream, we have $n_{\rm tracer}\propto n_{\rm true}$. This is a reasonable assumption since the particle spray algorithm we use has been proved to reproduce multiple morphological properties of GC streams with a typical error $\lesssim 10\%$ \citep{chen_improved_2025}. By defining $n_{\rm tracer}= \lambda\, n_{\rm true}$, we obtain 
\begin{equation*}
    N_{\rm tracer} = \lambda \iint n_{\rm true}(\phi_{1},\phi_{2})\, d\phi_1 d\phi_2 = \lambda\, N_{\rm true}.
\end{equation*}
Note that the ratio between the number of stars detected by \textit{StarStream} $N_{\rm detect}$ and $N_{\rm true}$ is the detection ratio $f_{\rm detect}$, a fundamental metric for stream detection methods. In \citetalias{chen_starstream_2025-1}, we have calculated $f_{\rm detect}$ for individual mock streams where $N_{\rm true}$ is known. Replacing $N_{\rm true}$ in the equation above with $f_{\rm detect}^{-1}N_{\rm detect}$, we obtain
\begin{equation}
    N_{\rm tracer} = \lambda\, f_{\rm detect}^{-1}N_{\rm detect}.
    \label{eq:N_tracer}
\end{equation}
We can also write $N_{\rm detect}$ as the integral of $n_{\rm true}(\phi_{1},\phi_{2})$ weighted by the spatial selection function $f_{\rm sel}(\phi_{1},\phi_{2})$
\begin{equation*}
    N_{\rm detect} 
    = \iint f_{\rm sel}(\phi_{1},\phi_{2})\, n_{\rm true}(\phi_{1},\phi_{2})\, d\phi_1 d\phi_2
    = \lambda^{-1}\iint f_{\rm sel}(\phi_{1},\phi_{2})\, n_{\rm tracer}(\phi_{1},\phi_{2})\, d\phi_1 d\phi_2.
\end{equation*}
Therefore, Eq.~(\ref{eq:N_tracer}) becomes
\begin{equation*}
   N_{\rm tracer} = f_{\rm detect}^{-1}\iint f_{\rm sel}(\phi_{1},\phi_{2})\, n_{\rm tracer}(\phi_{1},\phi_{2})\, d\phi_1 d\phi_2
\end{equation*}
where $\lambda$ and $\lambda^{-1}$ cancel. Plugging this equation back to Eq.~(\ref{eq:mass_loss_exact}) gives a simple formula for the mass loss rate:
\begin{equation*}
    \dot{M} \approx f_{\rm detect}^{-1}\frac{\dot{N}_{\rm tracer}\, M_{\rm sel}}{N_{\rm tracer}}.
\end{equation*}

\begin{figure}
    \centering
    \includegraphics[width=0.5\linewidth]{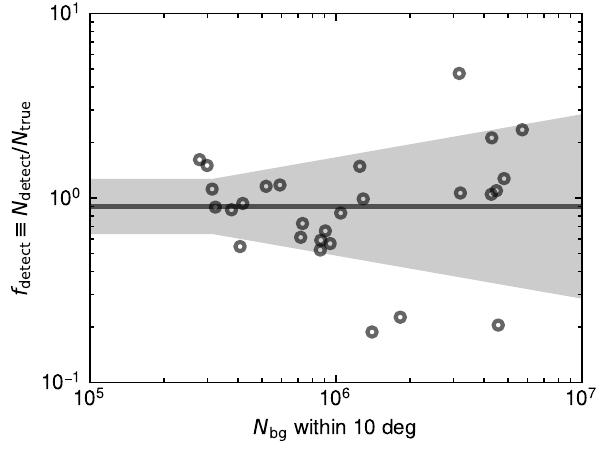}
    \caption{Detection ratio $f_{\rm detect}$ of \textit{StarStream} from \citetalias{chen_starstream_2025-1} as a function of the background density within $10^\circ$. Individual streams are shown as gray circles. The solid line stands for $f_{\rm detect,0}=0.9$, while the shaded ranges show the logarithmic scatter $\sigma_{\log f}$ parametrized as a function of $N_{\rm bg}$. Note that the data points here are selected with the same criteria as our high-quality sample, different from the \textit{upper panel} of Fig.~3 in \citetalias{chen_starstream_2025-1}, where no additional selection criterion is employed.}
    \label{fig:fdetect_vs_nbg}
\end{figure}

To approximate $f_{\rm detect}$ for individual streams in the high-quality sample, we use the test dataset in \citetalias{chen_starstream_2025-1} and apply the same selection criteria for this sample: $A_V<0.6$, $N_{\rm bg}<6\times10^6$, and $N_{\rm detect}>10$. In Fig.~\ref{fig:fdetect_vs_nbg}, we show $f_{\rm detect}$ as a function of the background density. We find that the $f_{\rm detect}$ is slightly below but consistent with unity in all ranges of $N_{\rm bg}$, while the scatter increases with $N_{\rm bg}$. This dependence can be well approximated by a constant mean of 0.9 with a log-normal scatter $\sigma_{\log f}$ increasing linearly from 0.15~dex to 0.5~dex when $\log_{10}N_{\rm bg}$ increases from 5.5 to 7. Since the test dataset lacks stream detections below $\log_{10}N_{\rm bg}<5.5$, we conservatively fix $\sigma_{\log f}=0.15$~dex for this region although the trend indicates a lower scatter. This simple parametrization is consistent with the more rigorous results by fitting a linear $\sigma_{\log f}$ -- $N_{\rm bg}$ relation using maximum likelihood estimation.

\section{Low-quality sample}
\label{sec:low_quality}

We provide the GC properties and number of detections $N_{\rm detect}$ for the low-quality sample in Table~\ref{tab:properties_low_quality}. Note that NGC 6715, Arp 2, Ter 7, and Ter 8 are members of the Sgr stream \citep[][marked with asterisk symbols in the table]{bellazzini_globular_2020,vasiliev_tango_2021}. We may overestimate $N_{\rm detect}$ by selecting stars from the Sgr stream, which shares similar spatial, proper motion, color, and magnitude distributions.

\begin{deluxetable}{lrrrrrrclrrrrrr}
\renewcommand\arraystretch{1.1} 
\tablecaption{Summary of GC and stream properties in the low-quality sample. The properties are described in Table~\ref{tab:properties}. A plain ASCII version is available at \url{https://github.com/ybillchen/StarStream_DR}.}
\label{tab:properties_low_quality}
\tablehead{
\colhead{GC} & \colhead{$l$} & \colhead{$b$} & \colhead{$d_\odot$} & \colhead{$A_V$} & \colhead{$N_{\rm bg}$} & \colhead{$N_{\rm detect}$} &
\colhead{} & 
\colhead{GC} & \colhead{$l$} & \colhead{$b$} & \colhead{$d_\odot$} & \colhead{$A_V$} & \colhead{$N_{\rm bg}$} & \colhead{$N_{\rm detect}$} \\
\colhead{} & \colhead{$(^\circ)$} & \colhead{$(^\circ)$} & \colhead{$({\rm kpc})$} & \colhead{} & \colhead{$(10^5)$} & \colhead{} &
\colhead{} & 
\colhead{} & \colhead{$(^\circ)$} & \colhead{$(^\circ)$} & \colhead{$({\rm kpc})$} & \colhead{} & \colhead{$(10^5)$} & \colhead{}
}
\startdata
Arp 2$^*$ & $8.5$ & $-20.8$ & $28.7$ & $0.31$ & $112.4$ & $8000$ &  & NGC 6402 & $21.3$ & $14.8$ & $9.1$ & $1.86$ & $77.0$ & $151$ \\
BH 140 & $303.2$ & $-4.3$ & $4.8$ & $2.11$ & $324.1$ & $20$ &  & NGC 6426 & $28.1$ & $16.2$ & $20.7$ & $1.12$ & $83.0$ & $117$ \\
Djor 2 & $2.8$ & $-2.5$ & $8.8$ & $2.91$ & $425.5$ & $14$ &  & NGC 6441 & $353.5$ & $-5.0$ & $12.7$ & $1.43$ & $461.0$ & $14$ \\
E 3 & $292.3$ & $-19.0$ & $7.9$ & $0.93$ & $46.0$ & $96$ &  & NGC 6496 & $348.0$ & $-10.0$ & $9.6$ & $0.74$ & $361.9$ & $865$ \\
IC 1257 & $16.5$ & $15.1$ & $26.6$ & $2.26$ & $123.2$ & $19$ &  & NGC 6535 & $27.2$ & $10.4$ & $6.4$ & $1.30$ & $71.7$ & $374$ \\
IC 4499 & $307.4$ & $-20.5$ & $18.9$ & $0.71$ & $51.5$ & $85$ &  & NGC 6541 & $349.3$ & $-11.2$ & $7.6$ & $0.34$ & $256.9$ & $777$ \\
NGC 2298 & $245.6$ & $-16.0$ & $9.8$ & $0.68$ & $42.9$ & $537$ &  & NGC 6544 & $5.8$ & $-2.2$ & $2.6$ & $2.36$ & $521.9$ & $351$ \\
NGC 2808 & $282.2$ & $-11.3$ & $10.1$ & $0.68$ & $109.9$ & $4438$ &  & NGC 6553 & $5.3$ & $-3.0$ & $5.3$ & $1.95$ & $498.1$ & $38$ \\
NGC 3201 & $277.2$ & $8.6$ & $4.7$ & $0.74$ & $85.7$ & $121$ &  & NGC 6569 & $0.5$ & $-6.7$ & $10.5$ & $1.64$ & $522.1$ & $154$ \\
NGC 4372 & $301.0$ & $-9.9$ & $5.7$ & $1.49$ & $207.7$ & $248$ &  & NGC 6584 & $342.1$ & $-16.4$ & $13.6$ & $0.31$ & $152.2$ & $200$ \\
NGC 4833 & $303.6$ & $-8.0$ & $6.5$ & $1.02$ & $195.9$ & $739$ &  & NGC 6624 & $2.8$ & $-7.9$ & $8.0$ & $0.87$ & $560.3$ & $843$ \\
NGC 5139 & $309.1$ & $15.0$ & $5.4$ & $0.37$ & $70.2$ & $920$ &  & NGC 6637 & $1.7$ & $-10.3$ & $8.9$ & $0.56$ & $494.7$ & $242$ \\
NGC 5286 & $311.6$ & $10.6$ & $11.1$ & $0.78$ & $103.1$ & $879$ &  & NGC 6638 & $7.9$ & $-7.2$ & $9.8$ & $1.27$ & $555.8$ & $15$ \\
NGC 5927 & $326.6$ & $4.9$ & $8.3$ & $1.30$ & $236.5$ & $11$ &  & NGC 6652 & $1.5$ & $-11.4$ & $9.5$ & $0.28$ & $413.7$ & $68$ \\
NGC 6101 & $317.7$ & $-15.8$ & $14.4$ & $0.16$ & $107.0$ & $128$ &  & NGC 6656 & $9.9$ & $-7.6$ & $3.3$ & $1.12$ & $474.7$ & $304$ \\
NGC 6121 & $351.0$ & $16.0$ & $1.9$ & $1.33$ & $157.9$ & $460$ &  & NGC 6715$^*$ & $5.6$ & $-14.1$ & $26.3$ & $0.46$ & $326.9$ & $20007$ \\
NGC 6139 & $342.4$ & $6.9$ & $10.0$ & $2.33$ & $221.7$ & $15$ &  & NGC 6723 & $0.1$ & $-17.3$ & $8.3$ & $0.16$ & $218.6$ & $45$ \\
NGC 6144 & $351.9$ & $15.7$ & $8.2$ & $1.36$ & $187.2$ & $17$ &  & NGC 6760 & $36.1$ & $-3.9$ & $8.4$ & $2.39$ & $176.8$ & $108$ \\
NGC 6171 & $3.4$ & $23.0$ & $5.6$ & $1.02$ & $46.9$ & $663$ &  & NGC 6779 & $62.7$ & $8.3$ & $10.4$ & $0.74$ & $132.7$ & $960$ \\
NGC 6254 & $15.1$ & $23.1$ & $5.1$ & $0.81$ & $28.4$ & $721$ &  & NGC 6809 & $8.8$ & $-23.3$ & $5.3$ & $0.59$ & $68.2$ & $1874$ \\
NGC 6287 & $0.1$ & $11.0$ & $7.9$ & $1.86$ & $312.3$ & $79$ &  & Pal 11 & $31.8$ & $-15.6$ & $14.0$ & $1.08$ & $131.8$ & $19$ \\
NGC 6304 & $355.8$ & $5.4$ & $6.2$ & $1.52$ & $310.0$ & $17$ &  & Pal 15 & $18.8$ & $24.3$ & $44.1$ & $1.24$ & $34.1$ & $24$ \\
NGC 6355 & $359.6$ & $5.4$ & $8.7$ & $2.39$ & $393.4$ & $19$ &  & Rup 106 & $300.9$ & $11.7$ & $20.7$ & $0.62$ & $125.2$ & $65$ \\
NGC 6356 & $6.7$ & $10.2$ & $15.7$ & $0.87$ & $221.3$ & $921$ &  & Ter 3 & $345.1$ & $9.2$ & $7.6$ & $2.26$ & $218.7$ & $29$ \\
NGC 6362 & $325.6$ & $-17.6$ & $7.7$ & $0.22$ & $98.8$ & $893$ &  & Ter 7$^*$ & $3.4$ & $-20.1$ & $24.3$ & $0.22$ & $140.5$ & $3312$ \\
NGC 6366 & $18.4$ & $16.0$ & $3.4$ & $2.20$ & $86.9$ & $731$ &  & Ter 8$^*$ & $5.8$ & $-24.6$ & $27.5$ & $0.37$ & $61.6$ & $10969$ \\
NGC 6397 & $338.2$ & $-12.0$ & $2.5$ & $0.53$ & $251.1$ & $202$ & \multicolumn{8}{r}{\footnotesize $^*$Members of the Sgr Stream.}
\vspace{2mm}
\enddata
\end{deluxetable}

\bibliography{references}
\bibliographystyle{aasjournalv7}

\end{document}